\documentclass[preprint,showpacs,preprintnumbers,amsmath,amssymb]{revtex4}

\usepackage{graphicx}
\usepackage{subfigure}
\begin{document}

\bibliographystyle{apsrev}
\title{The electronic structure of the antiferromagnetic semiconductor MnSb$_{2}$S$_{4 }$}
\author{S. F. Matar\footnote{Corresponding author: matar@icmcb-bordeaux.cnrs.fr}} 
\affiliation{Institut de Chimie de la Mati\`ere Condens\'ee de Bordeaux CNRS. Universit\'e Bordeaux 1.  
87 Avenue du Dr. Albert Schweitzer, F-33608 Pessac Cedex, France. }
\author{R. Weihrich, D. Kurowski, A. Pfitzner}
\affiliation{Institut f\"{u}r Anorganische Chemie, Universit\"{a}t Regensburg Universit\"{a}tsstra{\ss}e 31, D-93040 Regensburg, Germany.}
\author{V. Eyert}
\affiliation{Institut f\"ur Physik, Universit\"at Augsburg, 86135 Augsburg, Germany.}

\date{\today}

\begin{abstract}
The electronic band structures of orthorhombic ({\it oP28}) and monoclinic ({\it mC28}) MnSb$_{2}S_{4}$  were investigated with ab initio calculations in the local spin density approximation (LSDA) to the density functional theory (DFT). An analysis of the electronic properties and of the chemical bonding is provided using the augmented spherical wave (ASW) method considering nonmagnetic, ferromagnetic, ferrimagnetic and antiferromagnetic model orderings. In agreement with experimental results both modifications of MnSb$_{2}$S$_{4}$ are predicted to be antiferromagnetic. While the experimental band gap is missed for the monoclinic polymorph, the calculated band gap for orthorhombic MnSb$_{2}$S$_{4}$ is close to the experimental one. 
\end{abstract}

\maketitle
\section{Introduction}

Magnetic and semiconducting manganese sulphides attracted attention of solid 
state chemists since the early days of X-ray crystallography and magnetic 
structure investigations$^{1-5}$ done on haurite (MnS$_{2})$ and alabandite 
(MnS). They exhibit high magnetic moments due to the coordination of 
Mn$^{2+}$ in MnS$_{6}$ octahedra where it prefers a high spin state with 
five unpaired electrons. However, MnS$_{2}$ has been discussed as a rare 
example of a high spin to low spin transition under high pressure$^{6, 7}$. 

In the past few years the chemistry of magnetic manganese materials was 
enriched by fascinating discoveries mainly on multinary manganese oxides$^{6-8}$. 
Properties like the giant and colossal magnetoresistance (GMR, CMR) 
inspired new fields of research on magnetic semiconductors. Besides 
promising technological applications and experimental challenges there is an 
increasing demand and success of theoretical understanding of the underlying 
chemical bonding and electronic properties. The development and application 
of effective density functional (DFT) methods within the local spin density 
approximation (LSDA) still plays an increasing role herein$^{9-11}$. 

Fascinating properties were also discovered on manganese chalcogenides. MnS 
and MnS$_{2}$ show antiferromagnetic ordering while diluted magnetic semiconductors 
(DMS) based on MnS exhibit outstanding properties related to spintronic applications$^{12-14}$. 
Multinary materials like MnCr$_{2}$S$_{4}$ provide additional potential with respect to 
anisotropic resistivity and magnetic properties$^{15,16}$. 

Due to its reduced dimensionality MnSb$_{2}$S$_{4}$ serves as a promising 
low dimensional magnetic semiconducting material. Contrary to spinel type 
MnCr$_{2}$S$_{4}$ with Mn$^{2+}$ in MnS$_{4}$ tetrahedra one finds MnS$_{6}$ 
octahedra in MnSb$_{2}$S$_{4}$. Therein, it is related to MnS and MnS$_{2}$ as 
well as in the observation of phase transitions. Orthorhombic MnSb$_{2}$S$_{4}$ 
is accessible by hydrothermal synthesis and was earlier shown$^{17}$ to be isotypic to FeSb$_{2}$S$_{4}$ which is an 
antiferromagnetic material$^{18}$. Recently a new monoclinic modification 
(\textit{mC28}) of MnSb$_{2}$S$_{4}$ was synthesized by high temperature methods$^{19}$. 
MnSb$_{2}$S$_{4}$ (\textit{mC28}) can be transformed reversibly into the orthorhombic 
modification (\textit{oP28}) at high pressure$^{20}$. By electrical conductivity and 
magnetic susceptibility measurements it was found that MnSb$_{2}$S$_{4}$ (\textit{mC28}) is 
a semiconducting antiferromagnet with $T_{N}$ = 26.5 K and an electronic band 
gap of 0.77 eV$^{19,20}$. Concerning the bonding situation one faces 1D 
magnetic interactions, as well as bonds with and within the 
[$SbS_{3}]^{3-}$ ligand network that is related to Sb$_{2}$S$_{3}$ $^{24}$. 
However, no theoretical investigations are reported yet. Considering 
MnS and MnS$_{2}$ again as prominent examples, LSDA calculations$^{21-23}$ 
achieved good agreement with experimental results, {\it i.e.} the prediction of 
semiconducting and magnetic ground states with moments around 4.5 \textit{$\mu $}$_{B}$ for 
Mn$^{2+}$. For $\alpha $-MnS the antiferromagnetic ground state was correctly found$^{21,22}$. 
LSDA total energy calculations on MnS$_{2}$ supported the possibility of a 
low spin/high spin phase transition for a compressed cell$^{22}$. 

To discuss the differences and relations of the bonding, spin states and 
magnetic ordering in \textit{mC28} and \textit{oP28} MnSb$_{2}$S$_{4}$ first principles 
calculations are subsequently reported modelling non magnetic (NM), ferromagnetic (FM), 
ferrimagnetic (FIM for the monoclinic system) and antiferromagnetic (AFM) structures in order to identify the ground state configuration. The applied augmented spherical wave (ASW) method was successfully used in 
previous calculations on magnetic semiconducting manganites$^{11}$. The 
crystal structures, computational details, and results of the calculations 
on non-magnetic, ferromagnetic, and antiferromagnetic configurations are 
presented, as well as electronic band structures, site projected densities 
of states and chemical bonding characteristics.

\section{Crystal structures of orthorhombic and monoclinic phases}
For the calculations presented herein, the crystal structures of both 
MnSb$_{2}$S$_{4}$ modifications as determined by single crystal X-ray 
diffraction were taken as the starting points$^{17,19,20}$. The space groups and the 
relevant lattice parameters used in the calculation are given in the first part of Table 1. 
Both modifications are based on chains of edge-sharing MnS$_{6}$ octahedra (Fig. 1). These 
chains of octahedra are linked by [$SbS_{3}]^{3-}$ units to form 
layers in the case of MnSb$_{2}$S$_{4}$ (\textit{mC28}) and a three dimensional (3D) 
network in the case of MnSb$_{2}$S$_{4}$ (\textit{oP28}). 

The Sb-S bonds determine both the structural anisotropies and the 
differences between the modifications. Sb atoms exhibit a 3+2+$x$ ($x$ = 1, 2) 
coordination with three Sb-S bonds of about 2.5 {\AA} and two Sb-S bonds 
between 2.9 and 3.1 {\AA} (``secondary bonds''). In addition, there are 
so-called non bonding distances 3.1 {\AA}$<d$(Sb-S) $<$ 4 {\AA}. Distinguishing 
these three types of Sb-S interactions we find all short Sb-S bonds linking 
edge sharing MnS$_{6}$-octahedra of one chain in MnSb$_{2}$S$_{4}$ (\textit{mC28}). 
Slightly longer bonds link the octahedra to form a layered structure (Fig. 1). 
Between the layers (along the $c$ axis) only so-called non-bonding Sb-S 
distances are found. In the case of MnSb$_{2}$S$_{4}$ (\textit{oP28}) one finds double 
chains of octahedra which are interlinked by short and secondary Sb-S bonds. 
These double chains form a kind of fishbone scheme and non-bonding 
Sb-S distances between them result in a 3D network. The density of the title 
compound increases from 4.24 g/cm$^{3}$ (\textit{mC28})$^{19}$ to 4.51 g/cm$^{3}$ 
(\textit{oP28}) $^{17}$, showing that the orthorhombic modification is the high pressure 
form. The distances $d$(Mn-S) vary from around 2.6 \AA~  (Table 1) in both 
modifications. Thus, they show a slightly broader range than in the pure 
manganese sulphides with octahedral coordination of manganese, {\it i.e.}, 
$d$(Mn-S) = 2.61 {\AA} in $\alpha $-MnS $^{2}$ and $d$(Mn-S) = 2.59 {\AA} in 
MnS$_{2}$ $^{1}$. There are two different Mn positions in MnSb$_{2}$S$_{4}$ 
(\textit{mC28}), with a higher site symmetry than the single Mn position in 
MnSb$_{2}$S$_{4}$ (\textit{oP28}). The distortions of the MnS$_{6}$ octahedra are due to 
the structural anisotropy imposed by the [$SbS_{3}]^{3-}$ units. They result in tetragonally 
distorted MnS$_{6}$ octahedra with a coordination number of 2+4 in (\textit{mC28}) MnSb$_{2}$S$_{4}$ 
 and a coordination number of 1+1+2+2 in MnSb$_{2}$S$_{4}$ (\textit{oP28}), 
respectively. Further details are provided in refs.$^{17,19}$. 

Considering the magnetic coupling of manganese in the two polymorphs of 
MnSb$_{2}$S$_{4}$ the structural anisotropy provided by the MnS$_{6}$ chains 
has to be kept in mind. Thus, only two contacts $d$(Mn-Mn) $\approx $ 3.8 {\AA} 
are present in the title compound, and all other distances between Mn atoms 
are larger than 6 {\AA}. This situation is quite different from the isotropic Mn 
sublattices of, e.g., $\alpha $-MnS (12 x $d$(Mn-Mn) $\approx $ 3.7 {\AA}), and 
MnS$_{2}$ (12 x $d$(Mn-Mn) $\approx $ 4.3 {\AA}). This allows to investigate FM 
models with equal Mn spin directions and AFM models with alternating Mn 
spins along the chains in MnSb$_{2}$S$_{4}$.

\section{Computational framework}
The electronic properties have been self-consistently calculated in the 
framework of the density functional theory DFT$^{25,26}$ using the ASW method 
as implemented by Williams {\it et al.}$^{27}$ and Eyert$^{28}$. The effects of 
exchange and correlation were parameterized according to the local spin density 
approximation (LSDA) scheme of Vosko, Wilk and Nusair$^{29}$. 
All valence electrons, including 4$d$(Sb) ones, were treated as band states. 
In the minimum ASW basis set, we chose the outermost shells to represent 
the valence states and the matrix elements were constructed using partial waves 
up to $l_{max.}$ = 2 quantum number. The ASW method uses the atomic sphere approximation 
(ASA) which assumes overlapping spheres centered on the atomic sites where the 
potential has a spherical symmetry. In order to represent the correct shape of the 
crystal potential in the large voids of the respective crystal structures, additional 
augmentation spheres were inserted$^{28}$ to avoid an otherwise too large overlap 
between the actual atomic spheres. 

The calculations implicit of zero entropy ($T$ = 0 K) were started assuming a 
non-magnetic configuration which is non spin polarized (NSP) meaning that spin 
degeneracy was enforced for all species (atoms and empty spheres). Note that this configuration does not translate a paramagnetic state which 
would actually require a supercell with different orientations of the spins over the crystal sites. 
In a second step spin polarized (SP) calculations were performed by initially 
allowing for differing spin occupations, $i.e.$, majority (spin up $\uparrow $) 
and minority (spin down $\downarrow $) spins for all atomic species. 
The occupancies were self-consistently changed until convergence of the 
total energy ($\Delta$E$\leq$~10$^{-6}$ Ry.) and of the charges ($\Delta$Q$\leq$~10$^{-6}$) between two subsequent iterations was reached. For that a sufficiently large number of $k$ points was used with respect to selfconsistancy of the results. In view of the large cells especially when symmetry is broken by introducing the antiferromagnetic orderings, we used up to 12*12*12, {\it i.e.}\ 1728 points to produce respectively 216 and 468 {\it k} points in the irreducible wedges of the orthorhombic and monoclinic Brillouin zones. 
Calculations are implicit of collinear magnetic structures. However non-collinear magnetic structures can occur in manganese based compounds such as in the nitride Mn$_{4}$N which was studied in the same calculational framework$^{30}$. In fact such heavy calculations could be achieved with great accuracy in energy differences between the magnetic configurations provided one considers high symmetry structures such as that of cubic anti-perovskite Mn$_{4}$N. 
When one magnetic/crystallographic sublattice of all species 
is accounted for, a ferromagnetic order (FM) is described. Two magnetic 
sublattices need to be accounted for to calculate the AFM configurations. 
This can be achieved by symmetry breaking of the system, half of the 
constituents being ``spin up'' and the other half being ``spin down''. This 
approach accounts for the effect of low spin and high spin Mn$^{2+}$ and 
spin spin interactions in AFM and FM models for MnSb$_{2}$S$_{4}$ similar to 
the incommensurate magnetic structure of FeSb$_{2}$S$_{4}$$^{18}$. Indeed, 
spin reorientation, spin disorder, and the competition between AFM and FM 
orientation are discussed to play an important role in magnetic systems. We 
are aware of the fact that our models do not simulate spin dynamics. 
However, any spin spin interaction as for example in the incommensurate AFM 
structure of FeSb$_{2}$S$_{4}$ has to be expected between the states given 
by the AFM, FM and NM models.
Considering  the orthorhombic structure which has four MnSb$_{2}$S$_{4}$ formula units, two AFM configurations were accounted for, {\it i.e.} with the spin aligned oppositely in MnS$_{6}$ octahedral chains, this will be called hereafter AFM1 and another one with spins aligned parallel within a chain and oppositely between chains (AFM2). 
As for the monoclinic variety, the unit cell has two different manganese sites Mn1 and Mn2. This leads to a first possibility which is to account for antiparallel spin alignment between Mn1 and Mn2 sites leading to a ferrimagnetic (FIM) order. The other possibility is to double the cell along the third lattice vector c with Mn1 and Mn2 all up-$\uparrow$ in the first cell and Mn1 and Mn2 all down-$\downarrow$ in second cell, {\it i.e.} conforming with the spin spiral found  for MnSb$_{2}$S$_{4}$ (\textit{mC28})$^{31}$. Needless to say that  the symmetry breaking due to the magnetic lattice orderings among Mn($\uparrow$) and Mn($\downarrow$) in both structures computations are much heavier to carry out whence the limitation in the Brillouin zone integration in {\it k} points presented above.

Further  information about the nature of the interaction between atomic constituents can be provided using overlap population (OP) leading to the so-called COOP (crystal orbital overlap population)$^{32}$ or alternatively introducing the Hamiltonian based population COHP (crystal orbital Hamiltonian population)$^{33}$. Both approaches lead to a qualitative description of the chemical interactions between two atomic species by assigning a bonding, non-bonding or antibonding character.  A
slight refinement of the COHP was recently proposed in form of the ``covalent bond energy'' E$_{COV}$  which combines both COHP and COOP so as to make the resulting quantity independent of the choice of the zero of potential$^{34}$. The E$_{COV }$ was recently implemented within the ASW method$^{35}$. Our experience with both COOP and E$_{COV}$ shows that they give similar general trends although COOP exagerate the magnitude of antibonding states. We shall be using the E$_{COV}$ description of the chemical bonding.  

\section{Calculation results and discussion}
\subsection{Total energy and magnetic moments}
Charge transfer is observed from Mn towards Sb, S and the empty spheres; nonetheless its amount is not significant in terms of an ionic description (such as Mn$^{2+}$), which is rarely observed in the framework of such calculations. A more meaningful picture is provided from the quantum mixing of the valence states as it will be shown in the plots of the density of states (DOS) and the chemical bonding (E$_{COV}$) in next sections.
The two polymorphs show similar trends concerning the total energy 
calculated for the non magnetic (NM) and spin polarized (SP) ferromagnetic -FM-  and antiferromagnetic -AFM- models. Further ferrimagnetic calculations in the monoclinic system were carried out.  This is detailed in table 1 which presents the results obtained after self consistent computations for the different magnetic configurations considered. For both modifications the FM state is favored with respect to the non magnetic one. The large gain in energy arises from the magnetic exchange of coupled high spin Mn$^{2+}$ when spin polarization is accounted for. In FM configuration the resulting total magnetization per formula unit is close to 5 $\mu_B$. For formally Mn$^{2+}$ two configurations are possible for the spin arrangements within the octahedral field of sulphur: A high spin -HS- configuration $t_{2g}^{3}$, $e_{g}^{2}$ with 5 unpaired spins and a low spin -LS- one: $t_{2g}^{5}$, $e_{g}^{0}$ resulting in only one unpaired spin. From this it can be suggested that at least in the ferromagnetic state divalent manganese is HS. However the total magnetization arises from all constituents of the lattice (Table 1) with the main contribution due to manganese. The SP-FM Mn moment amounts to 4.38 $\mu_B$ within (\textit{oP28}) MnSb$_{2}$S$_{4}$ and 4.23 \textit{$\mu$}$_{B}$ for (\textit{mC28}) MnSb$_{2}$S$_{4}$. This agrees with the value calculated for Mn$^{2+}$ in MnTe$^{21}$ and lies in the range of further LSDA results on high spin Mn$^{2+}$ from 4.3 to 4.6  \textit{$\mu $}$_{B}$$^{21-23}$. When the ferrimagnetic configuration is accounted for within the monoclinic variety a further stabilization is obtained and there is a cancelling out between moments so that  total magnetization is zero. The AFM configurations show a further energy lowering for both varieties with a smaller energy difference in the orthorhombic structure. This would suggest a lower N\'eel temperature for the high pressure orthorhombic variety. Within this structure the AFM1 configuration with the spins aligned oppositely in MnS$_{6}$ octahedral chains is found to be favored with respect to the AFM2 one ({\it i.e.} with spins aligned parallel within a chain and oppositely between chains). The small lowering in the moment carried by Mn$^{2+}$ (4.34 \textit{$\mu $}$_{B})$ agrees with the value observed for $\alpha $-MnS (4.1 \textit{$\mu $}$_{B}$)$^{22}$. For MnSb$_{2}$S$_{4}$ (\textit{mC28}) we find the same order of energies. Thus, the applied method predicts the preference of an antiparallel coupling of the spins of Mn-$d$ electrons in a high spin state against a parallel coupling. The close magnitudes of the moments between the FM and the AFM configurations lead to propose that the magnetic order might be Heisenberg-like whereby the magnetic susceptibility should obey a Bonner-Fischer behavior$^{36}$ which is characteristic to linear spin chains. 

\subsection{Non-spin polarized calculation DOS and chemical bonding}
The suggested NSP situation for MnSb$_{2}$S$_{4}$ (\textit{oP28}) and (\textit{mC28}) results in a metallic behavior, analogous to studies on MnS and MnS$_{2}$$^{20-22}$. The site projected DOS are shown in fig. 2 a and b. The highest occupied states cross the Fermi level $E_{F}$ at a high density of states which is attributed to t$_{2g}$ states from a crystal field analysis of Mn $d$ states projections given in fig. 2c. These Mn$^{2+}$ $t_{2g}$ states are only partly occupied by five electrons. The next bands above $E_{F}$ are formed by the Mn $e_{g}$-states (fig. 2c). The splitting of the $e_{g}$ states results from deviations of the MnS$_{6}$-groups from octahedral symmetry. Antimony and sulphur $p$-states form broad bonding states with the metal states in the energy range [-6,- 1eV] (see next paragraph). The DOS at low energies are $s$-bands of Sb (-10 eV) and S (-15 eV); the latter are found at lower energy due to the higher electronegativity of sulphur as compared to antimony. 

The chemical bonding within both orthorhombic and monoclinic MnSb$_{2}$S$_{4}$ are examined in the framework of the E$_{COV}$$^{34,35}$ for Mn-S, Mn-Sb and Sb-S pair interactions. The corresponding covalent bond energy E$_{COV}$ plots are given in fig. 3 a and b. Negative, positive and nil E$_{COV}$ magnitudes are relevant to bonding, antibonding and non-bonding characteristics. From this the major part of the valence band VB is bonding due to Mn-S interactions as well as to Sb-S albeit with a smaller magnitude; this contributes to the stabilization of the crystal lattice. The Sb-S interaction is observed with smaller magnitude and it remains bonding within the conduction band above E$_{F}$. This somehow provides an illustration for the description of the bonding given in the crystal structure section above. Mn-Sb interaction plays little role -as with respect to Mn-S one- within the major range of the VB.  At the top of the VB the system becomes largely destabilized as the Fermi level is reached, {\it i.e.} where  a large Mn-S E$_{COV}$ as well as Mn-Sb antibonding interactions with smaller magnitude can be observed.  Although a large part of the Mn(t$_{2g}$) are not engaged into Mn-S antibonding interaction in asfar as they are responsible for the onset of the Mn magnetic moment, the non magnetic configuration is clearly not favored from that. Lastly Mn-Mn interaction were observed too but with a much smaller magnitudes than all other explicited ones in both crystal varieties, so they are not shown here. Nevertheless it will be discussed below that these bonds can have consequences on the electronic structure (cf. section III.C particularly for the monoclinic band structures).

\subsection{The electronic structure of spin polarized MnSb$_{2}$S$_{4}$ }
\subsubsection{Ferromagnetic state}
As shown by the site projected DOS in figs. 4  a  and  b, the spin polarization causes Mn 3$d$ levels to split into majority spin ($\uparrow $) states which are lowered in energy relative to minority spin ($\downarrow $) states at higher energy. Majority Mn $d$ spin states completely lie below $E_{F}$, thus being fully occupied by 5 electrons. The minority Mn $d$ states are found above $E_{F}$ thus being completely empty. This indicates a closely non metallic situation with a small energy  gap in the orthorhombic variety which reduces to a closing in (\textit{mC28}) MnSb$_{2}$S$_{4}$. The DOS for manganese in both varieties exhibit peaks which closely resemble the $t_{2g}(\uparrow$)-$e_{g}(\uparrow $) manifolds. Thus, the highest occupied states in the valence band are formed by Mn up spin $e_{g}$ states and the lowest unoccupied ones by down spin $t_{2g}$ states. Concerning Sb and S DOS the latter can be observed to closely follow the shape of Mn pointing to the Mn-S coordination, {\it i.e.} with  MnS$_{6}$ octahedra within which the major part of the bonding within the lattice occurs as discussed above. Spin polarization mainly affects Mn states so that there is hardly any energy shift between ($\uparrow $) and ($\downarrow $) spin populations for Sb and S although residual moments were computed in both orthorhombic and monoclinic systems (Table 1). 

\subsubsection{Ferrimagnetic (FIM) model in  MnSb$_{2}$S$_{4}$(\textit{mC28}) } A first possibility to account for antiparallel spin alignment within (\textit{mC28}) MnSb$_{2}$S$_{4}$ was to allow for it between the two singly occupied Mn sublattices within the base centered monoclinic structure. The resulting energy differences shown in Table 1 are found in favor of this FIM configuration by 13.3 meV with respect to FM. The magnitudes of the moments are within range of FM calculations but the resulting magnetization is zero. The DOS and band structure given in fig. 5 a and b respectively. The DOS plot  shows some similar features to FM (fig. 4b)  but there is now a gap opening  in the minority spins whereas a metallic behaviour is observed for majority spins. From the band structure plot in the same energy window the gap of $\sim$0.6 eV can be observed between the VB and the CB  in U(A-E) direction which is along the $k_z$ axis of the monoclinic Brillouin zone. It is along this direction that the metallic behavior is obtained too as resulting from  the crossing of single bands from the VB and the CB due to a large dispersion. Thus the monoclinic system, in an intermediate magnetic state (see relative energies in Table 1), is not a semiconductor but a half-metallic ferrimagnet with a relatively low DOS at E$_{F}$ due to single band crossing. 

\subsubsection{Antiferromagnetic (AFM) models} 
For all systems the energy differences shown in Table 1 are in favor of AFM ground state configurations (AFM1 for \textit{oP28}). The  result of enforced AF configuration is that the total up spin and down spin projected densities of states present the same contributions. As a consequence plots for one magnetic sublattice within each structure will be shown. In a narrow energy window around the Fermi level meant to exhibit the relevant features of the AFM ground state, figs. 6 and  7  give the DOS and band structure for orthorhombic and monoclinic AFM MnSb$_{2}$S$_{4}$ respectively. 
The MnSb$_{2}$S$_{4}$ (\textit{oP28}) projected DOS (fig. 6) show a larger splitting around E$_{F}$ than in the FM DOS (fig. 4a). The larger gap is likely to arise from a shift of unoccupied minority Mn states to higher energies within the CB which can be a result of Mn-Mn interactions throughout the MnS$_{6}$ chains.  From fig. 6b showing the band structure its magnitude amounts to $\sim$~0.7 eV between $\Gamma$$_{VB}$ and $\Gamma$$_{CB}$ for instance in the orthorhombic Brillouin zone.  This results in a nonconducting state. Note that this gap for the AFM state is close to the experimental value of 0.77 eV$^{20}$.  Our calculations indicate the preference of an AFM configuration (AFM1, cf. energy differences in table 1) based on a simple model of alternating Mn moments along the rods. This is somehow similar to the $\alpha $-MnS case examined by Tappero {\it et al.}$^{23}$. 

AFM ground state site projected DOS of  MnSb$_{2}$S$_{4}$(\textit{mC28}) (fig. 7) show different features from the ferrimagnetic case (fig. 5a) because both Mn1 and Mn2 are now polarized up or down within a magnetic sublattice (see for instance the change of orientation of Mn1 and Mn2 DOS above E$_{F}$); this results in larger $n(E_{F})$. In termes of band structure (fig. 7b) this involves enhanced band crossing along AE direction (along $k_z$ direction) as it can be observed from the confrontation with the ferrimagnetic band structure (fig. 5b). From such a band dispersion and crossing the system is obtained as weakly metallic. This is somehow opposed to the semiconducting state proposed experimentally. Nevertheless both monoclinic and orthorhombic varieties have been shown to possess similar features and the final answer on the question for the coupling of the magnetic moments will be given by neutron diffraction. Related investigations are in progress$^{31}$. 

\section{Conclusion}

The electronic structure of MnSb$_{2}$S$_{4}$ in both, the orthorhombic 
and the monoclinic modifications were calculated within the 
local spin approximation for non magnetic as well as for spinpolarized  
ferromagnetic, ferrimagnetic and antiferromagnetic models. 
According to total energy calculations the spin polarized states with high 
spin Mn$^{2+}$ are largely preferred to a non spin polarized one (Table 1). 
Magnetic moments of $\sim$4.3 \textit{$\mu $}$_{B}$ are calculated in 
agreement with high spin Mn$^{2+}$ configuration known from MnS and 
MnS$_{2}$. For both MnSb$_{2}$S$_{4}$ varieties the AFM model shows 
an additional energy gain, thus becoming the ground state. These results 
are accompanied by significant differences in the electronic structures 
of the models. The NSP model leads to a metallic behavior for both modifications 
with a partly filled VB formed by Mn $t_{2g}$ and the CB by the empty Mn $e_{g}$ 
bands shown by a crystal field analysis. In the orthorhombic system FM and AFM 
models lead to the experimentally observed semiconducting characteristics 
with a larger gap obtained for the AFM ground state. Differences in the electronic 
structures concerning the CB and the VB are due to the crystal structures. 
Calculations for MnSb$_{2}$S$_{4}$ (\textit{oP28}) reveal a  
band gap of 0.7 eV, close to the experimental value of 0.77 eV. 
In  MnSb$_{2}$S$_{4}$(\textit{mC28})  two Mn sites are 
present which have a significantly different environment by sulphur and 
therefore the site projected DOS for Mn shows a broadening, hence the VB is 
broadened too in comparison to the orthorhombic modification. On the other hand, 
the empty minority spin Mn $d$ states in the conduction band are sharper for 
the monoclinic modification. This is related to the higher local symmetry at 
the Mn sites. The computed intermediate ferrimagnetic state exhibits a half metallic 
behavior due to single Mn bands crossing along the AE direction in the Brillouin zone, 
{\it i.e. } along $k_z$. This is enhanced in the  AFM ground state. Although the 
antiferromagnetic nature of the ground state of both 
modifications of MnSb$_{2}$S$_{4}$ becomes evident by the present 
calculations, further investigations of electrical conductivity to reveal the semiconducting 
properties are needed, they are underway. 

\begin{acknowledgments}
Computational facilities were provided within the intensive numerical simulation
facilities network M3PEC of the University Bordeaux 1, partly financed by
the {\it Conseil R\'egional d'Aquitaine}. 
Support from the Deutsche 
Forschungsgemeinschaft (DFG) through Sonderforschungsbereich 484 is equally acknowledged.
\end{acknowledgments}

{}
\newpage
\begin{table}
\begin{tabular}{lcccl}
Parameters$^{19,20}$&		MnSb$_{2}$S$_{4}$ (\textit{oP}28)& MnSb$_{2}$S$_{4}$ (\textit{mC}28)\\
\hline
Space group& Pnam (62)&    		C2/m (12) \\
a b c (\AA)&   11.457 14.351 3.823 &	12.747 3.799 15.106 $\beta$=113.9$^{o}$ \\
$<d_{Mn-S} (\AA)>$&                2.588& 		2.611\\
$<d_{Sb-S} (\AA)>$&               2.551&		2.532\\
\hline
$\Delta$E$_{FM-NSP}$(eV/fu)&  -1.441 & 	-1.436	\\
$\Delta$E$_{FIM-FM}$(eV/fu)&  & 	-0.0133& \\
$\Delta$E$_{AFM-FM}$(eV/fu)& -0.020 & -0.073	 \\
$\Delta$E$_{AFM1-FM}$(eV/fu)& -0.014 & 	\\
\hline
M$^{FM}_{Mn}$ ($\mu_B$)& 	     4.384& 		4.138 / 4.200 \\
M$^{FM}_{S}$ ($\mu_B$)& 	    0.060 / 0.110& 		0.059 /  0.060\\
M$^{FM}_{Sb}$ ($\mu_B$)& 	    0.060 /  0.080& 		0.08 /  0.13\\
M$^{FM}_{cell}$ ($\mu_B$)& 	    20.0& 		9.82 \\
\hline
M$^{FIM}_{Mn}$ ($\mu_B$)&   & 		 +4.180  / -4.11\\
M$^{FIM}_{S}$ ($\mu_B$)&     & 		+0.049 /  -0.047 \\
M$^{FIM}_{Sb}$ ($\mu_B$)&    & 	-0.116 / +0.095 / +0.055 /  -0.088  \\
M$^{FIM}_{cell}$ ($\mu_B$)&   & 			0 	\\
\hline
M$^{AFM1}_{Mn}$ ($\mu_B$)& 	     $\pm$4.341& $\pm$(4.181/ 4.121) \\
M$^{AFM1}_{S}$ ($\mu_B$)& 	    $\pm$0.0003 /  0.0& $\pm$(0.001 / 0.09)\\
M$^{AFM1}_{Sb}$ ($\mu_B$)& 	    $\pm$0.056 &  $\pm$(0.004 / 0.052) \\
M$^{Spin~\uparrow - Spin~\downarrow}_{cell}$ ($\mu_B$)&$\pm$9.1& $\pm$8.93 \\
M$^{AFM1}_{cell}$ ($\mu_B$)&	     0&			 0	\\
\hline
M$^{AFM2}_{Mn}$ ($\mu_B$)& 	     $\pm$4.35& 		\\
M$^{AFM2}_{S}$ ($\mu_B$)& 	    $\pm$0.051 / $\pm$0.019& 	 \\
M$^{AFM2}_{Sb}$ ($\mu_B$)& 	    $\pm$0.092 / $\pm$0.029& 	 \\
M$^{Spin~\uparrow - Spin~\downarrow}_{cell}$ ($\mu_B$)& $\pm$9.00& 	 \\
M$^{AFM2}_{cell}$ ($\mu_B$)& 	     0 			&		\\
\hline
\end{tabular}
\label{tab1}

\caption{\label{tab:table1}Crystal data from literature and calculation results for orthorhombic (\textit{oP28}) and monoclinic (\textit{mC28}) MnSb$_{2}$S$_{4}$ -NSP=non spin polarized; FM=ferromagnetic; FIM=ferrimagnetic; AFM= antiferromagnetic.}
\end{table}
\newpage
\begin{figure}[htbp]
\subfigure[~]{\includegraphics[width=0.67\linewidth]{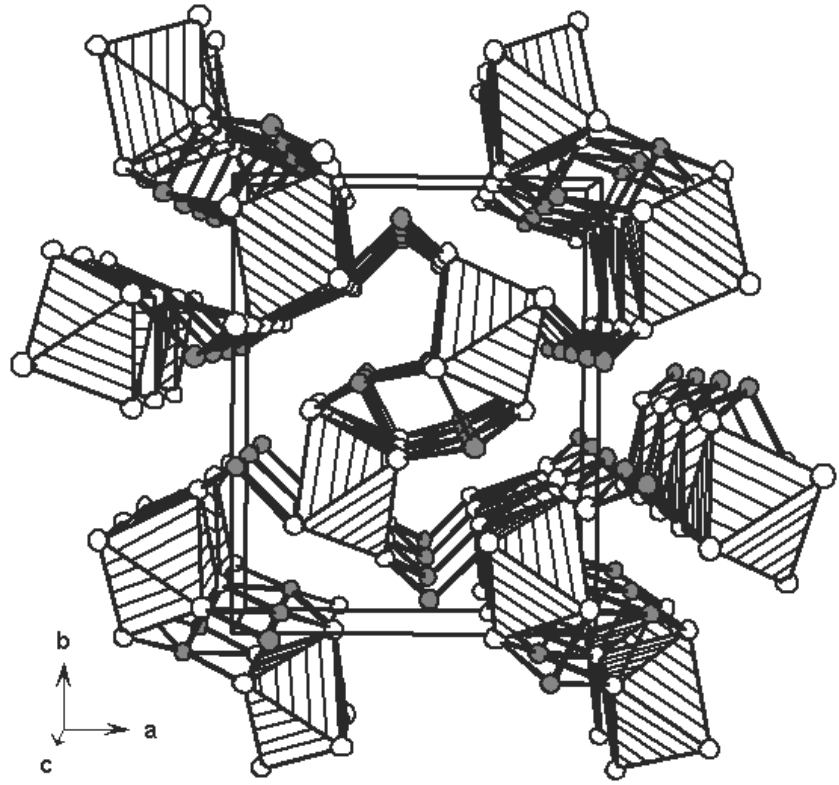}}
\subfigure[~]{\includegraphics[width=0.67\linewidth]{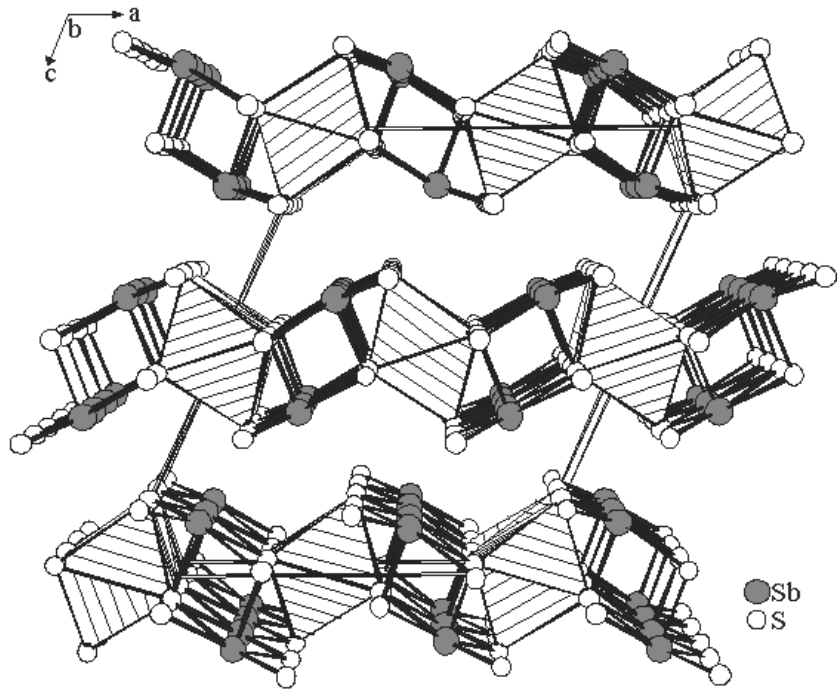}}
\label{fig1}
\caption{Crystal structures of a) orthorhombic and b) monoclinic 
MnSb$_{2}$S$_{4}$. View along the chains of edge sharing MnS$_{6}$ 
octahedra, S atoms are white, Sb grey. Bonds between Sb and S are drawn only 
for d(Sb-S) $<$ 3.15 {\AA}}
\end{figure}

\begin{figure}[htbp]
\begin{center}
\subfigure[~]{\includegraphics[width=0.5\linewidth]{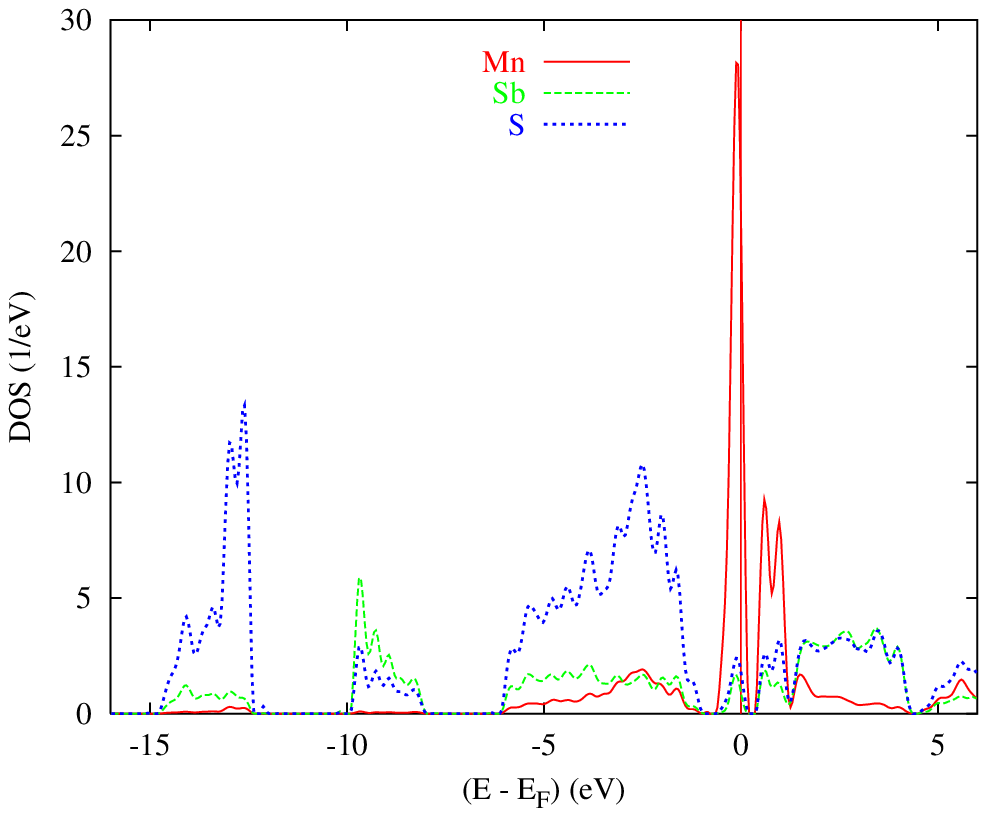}}
\subfigure[~]{\includegraphics[width=0.5\linewidth]{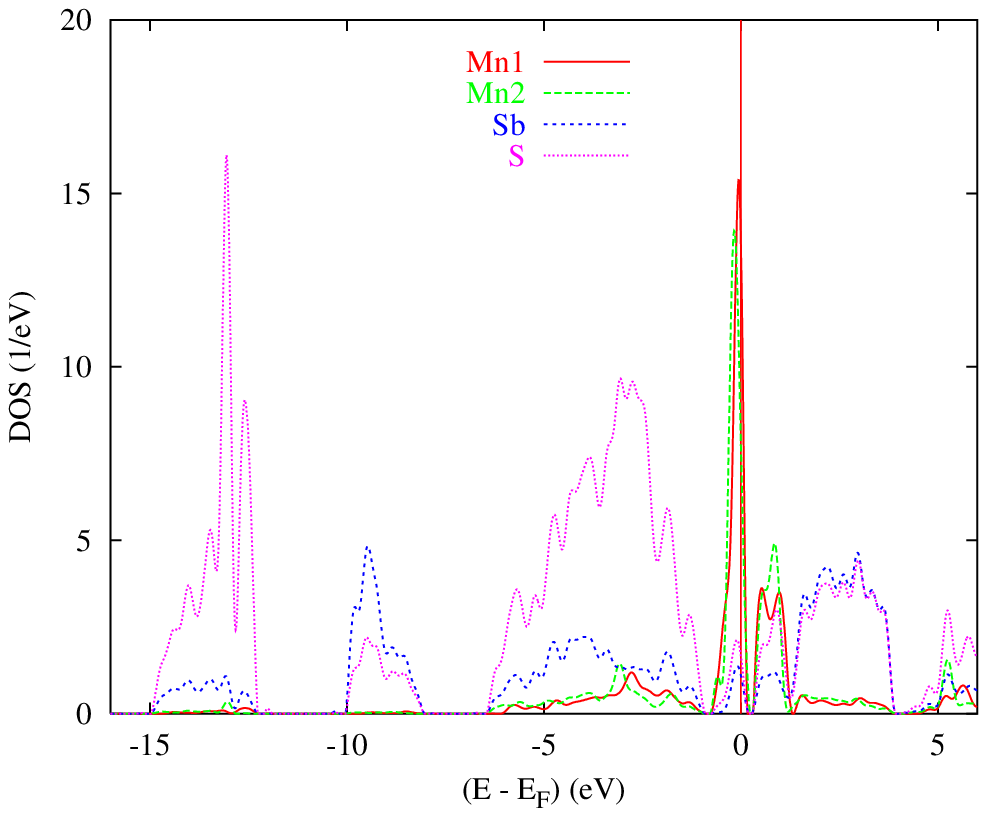}}
\subfigure[~]{\includegraphics[width=0.6\linewidth]{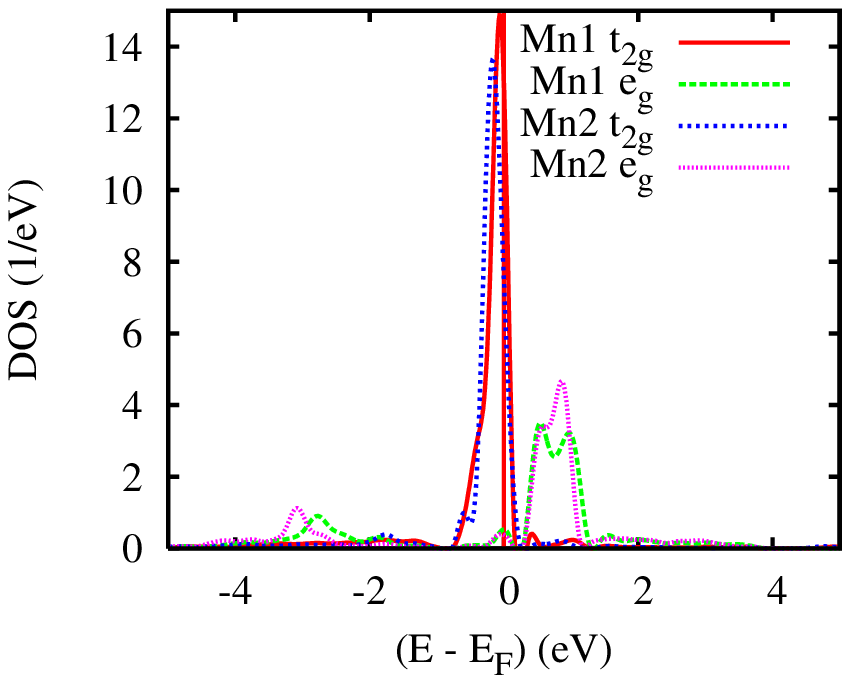}}
\caption{a) and b) Show the site projected DOS for one formula unit of non magnetic MnSb$_{2}$S$_{4}$ (resp. \textit{oP28} and \textit{mC28}). c) O$_{h}$ crystal field splitting of Mn1 and Mn2 sites in  MnSb$_{2}$S$_{4}$(\textit{mC28}). }
\label{fig2}
\end{center}
\end{figure}

\begin{figure}[htbp]
\begin{center}
\subfigure[~]{\includegraphics[width=0.67\linewidth]{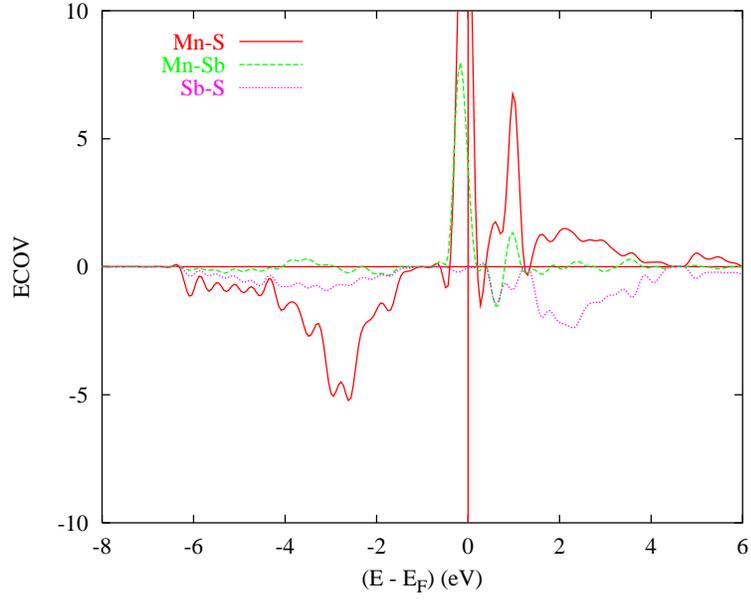}}
\subfigure[~]{\includegraphics[width=0.67\linewidth]{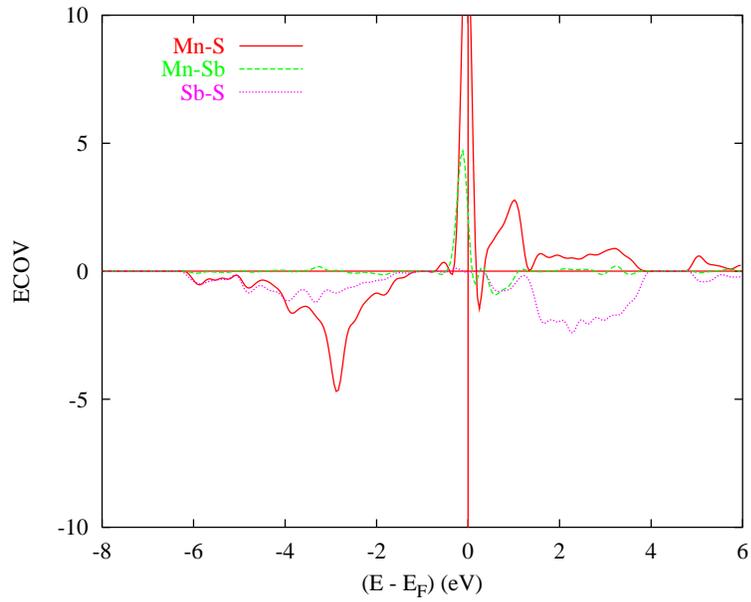}}
\caption{Chemical bonding properties from covalent bond energy E$_{COV}$ approach within MnSb$_{2}$S$_{4}$ per formula unit: a) \textit{oP28} orthorhombic variety, b) \textit{mC28} monoclinic variety for one of the two manganese sites -Sb and S regroup partial contributions from all lattice sites.}
\label{fig3}
\end{center}
\end{figure}

\begin{figure}[htbp]
\begin{center}

\subfigure[~]{\includegraphics[width=0.67\linewidth]{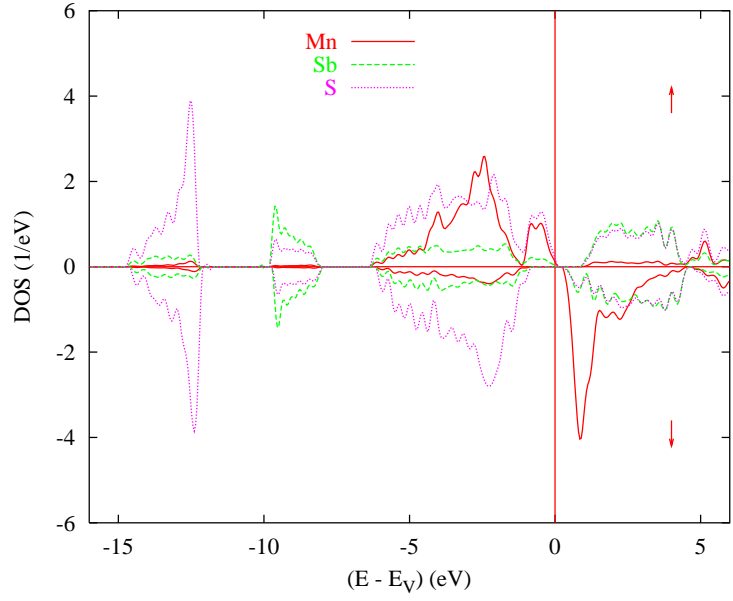}}
\subfigure[~]{\includegraphics[width=0.67\linewidth]{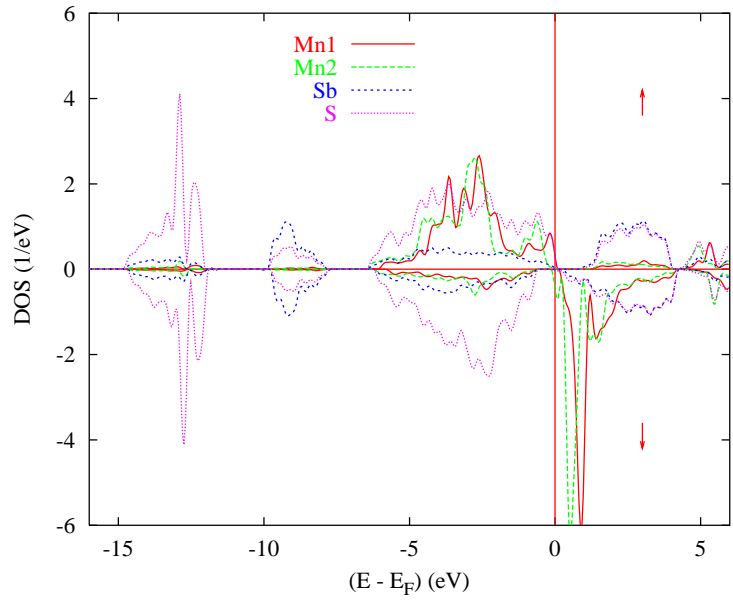}}

\caption{Spin resolved site projected DOS per formula unit for a)   
MnSb$_{2}$S$_{4 }$(\textit{oP28}), b) MnSb$_{2}$S$_{4}$ (\textit{mC28}) -Sb and S regroup partial contributions from all lattice sites.}
\label{fig4}
\end{center}
\end{figure}

\begin{figure}[htbp]
\begin{center}
\subfigure[~]{\includegraphics[width=0.6\linewidth]{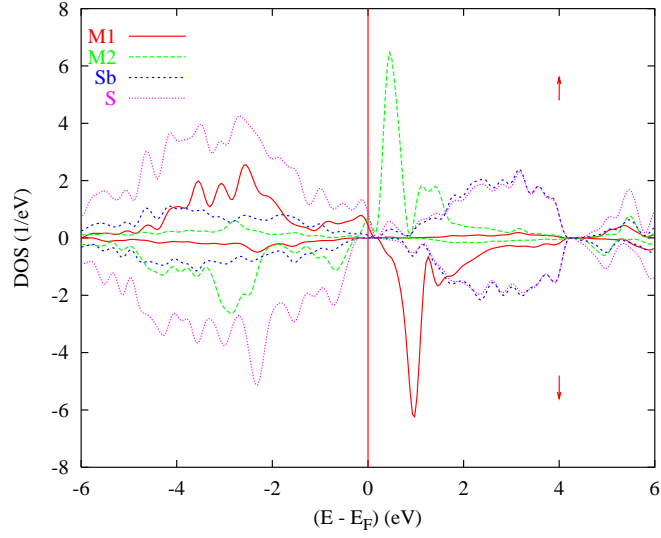}}
\subfigure[~]{\includegraphics[width=0.7\linewidth]{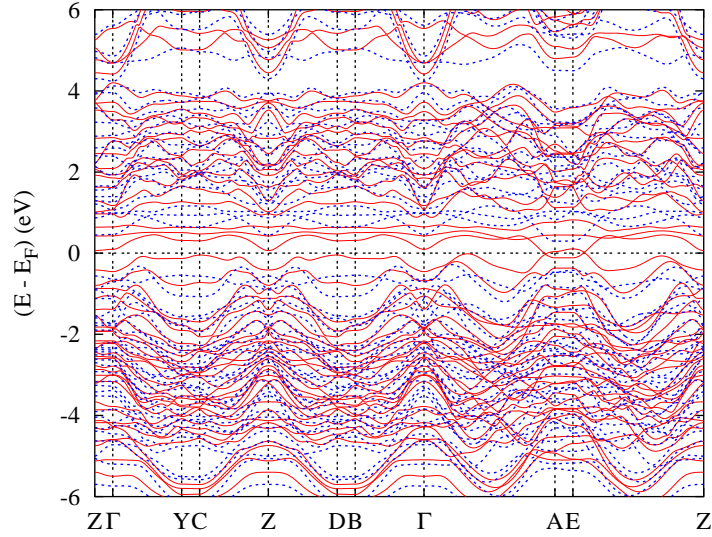}}
\caption{a) DOS (M1 and M2 stand for Mn1 and Mn2 respectively;  Sb and S regroup partial contributions from all lattice sites) and b) band structure in a narrow energy window around the Fermi level of ferrimagnetic intermediate state of monoclinic MnSb$_{2}$S$_{4}$(\textit{mC28}) (Solid lines ($\uparrow$), dotted lines ($\downarrow$)).}
\label{fig5}
\end{center}
\end{figure}

\begin{figure}[htbp]
\begin{center}
\subfigure[~]{\includegraphics[width=0.67\linewidth]{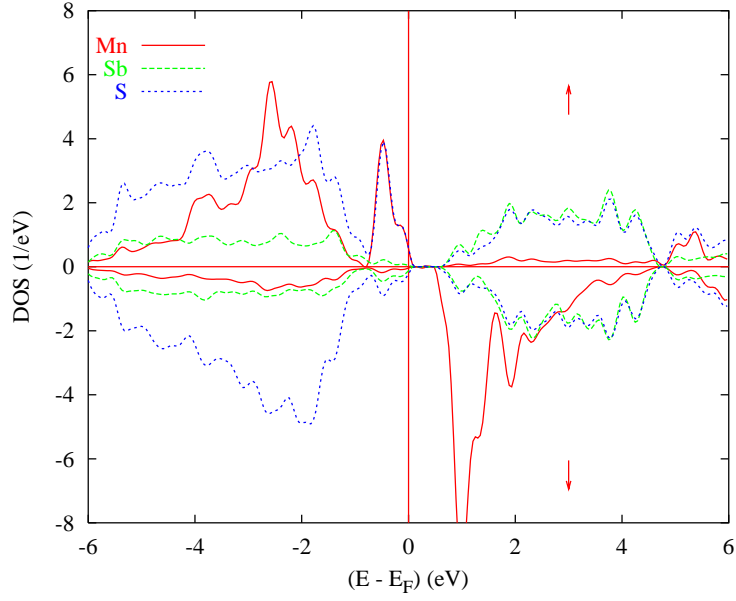}}
\subfigure[~]{\includegraphics[width=0.67\linewidth]{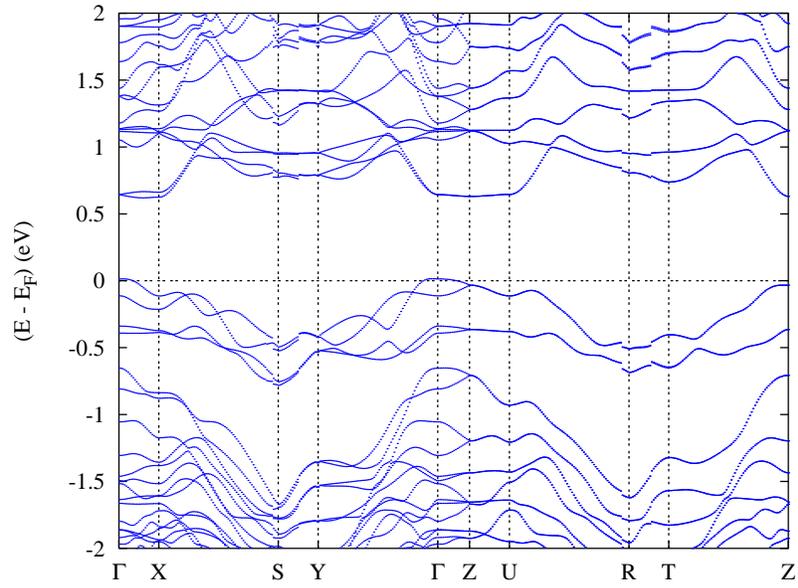}}
\caption{DOS and band structure in a narrow energy window around E$_{F}$ of antiferromagnetic ground state of orthorhombic  MnSb$_{2}$S$_{4}$(\textit{oP28}).}
\label{fig6}
\end{center}
\end{figure}

\begin{figure}[htbp]
\begin{center}
\subfigure[~]{\includegraphics[width=0.67\linewidth]{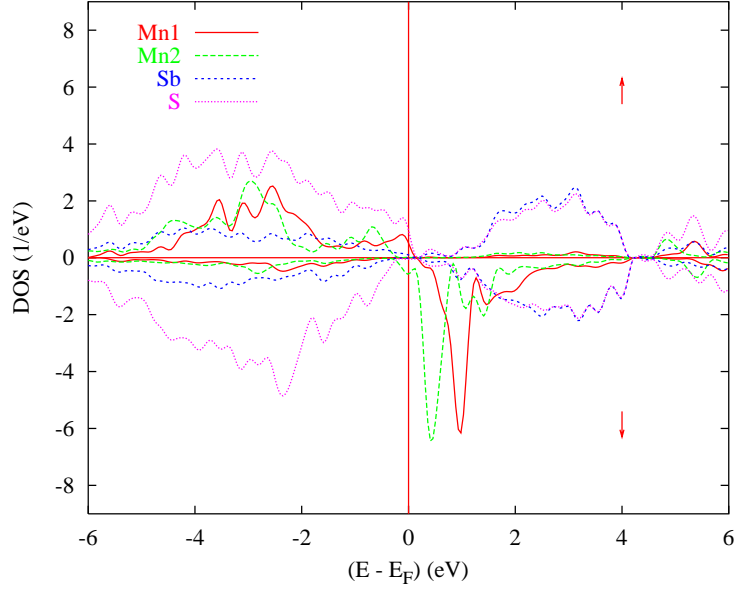}}
\subfigure[~]{\includegraphics[width=0.67\linewidth]{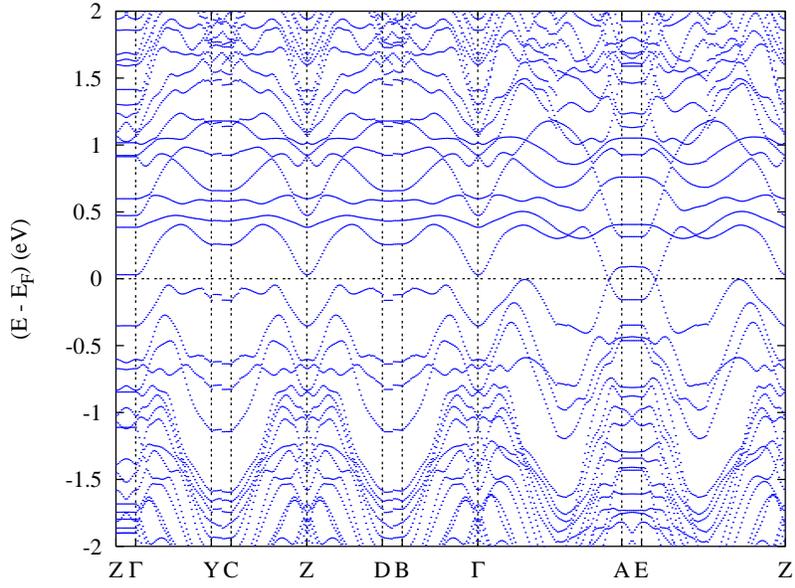}}
\caption{DOS and band structure  in a narrow energy window around E$_{F}$ of antiferromagnetic ground state of monoclinic  MnSb$_{2}$S$_{4}$(\textit{mC28}).}
\label{fig7}
\end{center}
\end{figure}

\end{document}